\documentclass[twocolumn,showpacs,amsmath,aps]{revtex4}
\usepackage{graphicx,color}
\usepackage{CJK}
\usepackage{bm}
\usepackage[hypertex]{hyperref}
\usepackage{float}

\newcommand{\be}{\begin{equation}}
\newcommand{\ee}{\end{equation}}
\newcommand{\bea}{\begin{eqnarray}}
\newcommand{\eea}{\end{eqnarray}}
\newcommand{\bsube}{\begin{subequations}}
\newcommand{\esube}{\end{subequations}}

\newcommand{\Eq}[1]{Eq.\,(\ref{#1})}
\newcommand{\Eqs}[1]{Eqs.\,(\ref{#1})}

\newcommand{\ra}{\rangle}

\newcommand{\nl}{\nonumber \\}

\newcommand{\beq}{\begin{equation}}
\newcommand{\eeq}{\end{equation}}
\newcommand{\beqn}{\begin{eqnarray}}
\newcommand{\eeqn}{\end{eqnarray}}
\newcommand{\bsub}{\begin{subequations}}
\newcommand{\esub}{\end{subequations}}

\begin{document}

\begin{CJK*}{GBK}{Song}

\title{Quantum transfer through a continuum under continuous monitoring}

\author{Luting Xu}
\email{xuluting@tju.edu.cn}
\affiliation{Center for Joint Quantum Studies and Department of Physics,
School of Science, Tianjin University, Tianjin 300072, China}

\author{Xin-Qi Li}
\email{xinqi.li@tju.edu.cn}
\affiliation{Center for Joint Quantum Studies and Department of Physics,
School of Science, Tianjin University, Tianjin 300072, China}


\date{\today}

\begin{abstract}
In this work we extend our previous studies on the quantum transfer of a particle
through a finite-bandwidth continuum under frequent detections,
by replacing the assumed frequent measurements with
a genuine continuous monitoring by a point-contact detector.
We present a quantitative comparison between the two types of measurement.
We also propose possible measurements, based on the state-of-the-art experiments,
to test the `scaling' property between the measurement rate
and the bandwidth of the reservoir,
rooted in the transfer dynamics under continuous monitoring.
\end{abstract}

\maketitle

\section{Introduction}

In a series of recent studies \cite{SG11,SG13,SG14,Xu16,Xu18,Xu19},
the `null'-result-conditioned dynamics of electron transfer through a continuum
or spontaneous emission of photons under continuous monitoring was analyzed.
In these studies the continuous monitoring in the reservoir has been considered as
a series of $\tau$-interval instantaneously projective measurements,
i.e., frequent checks if the electron/photon is in the reservoir or not,
after every time interval $\tau$.
For the Markovian case (wide-band limit),
a {\it no-effect} of measurements on the dynamics was concluded \cite{SG11,SG13}.
However, for the non-Markovian reservoir
(with finite bandwidth) \cite{SG14,Xu16,Xu18,Xu19},
the dynamics will be drastically influenced by
the time interval $\tau$ between successive measurements,
more specifically, being governed by a scaling parameter
$x=\Lambda\tau$, where $\Lambda$ is the spectral bandwidth of the reservoir.
Actually, this $x=\Lambda\tau$-scaling property has somehow extended
the well known quantum trajectory (QT) theory \cite{Dali92,WM93,WM09,Jac14}
-- which was constructed under continuous monitoring in Markovian environments --
to the case of non-Markovian environments.
The $x=\Lambda\tau$-scaling behaviors also establish a simple connection
between the QT theory and the quantum Zeno effect \cite{Kur00}.

We notice that the concept of continuous measurements has also been employed
as a theoretical tool (but not real measurements performed)
to analyze the effects of environment,
e.g., the environment induced decoherence and the appearance of classical feature
from a full quantum world \cite{Cav87,Bar05,Men03,Bar06}.
However, rather than taking as a theoretical tool,
for a real measurement-conditioned evolution,
it should be very hard to implement the frequent projective measurements
(with the short time interval $\tau$) employed in the theoretical considerations. 
For photon detections, the time interval $\tau$ may roughly correspond to
the signal-response-time of the detector.
However, in the set-up of electron transfer through a reservoir,
how to implement the frequent projective measurements
in the reservoir is rather unclear.
In this work, instead of the projective measurements introduced in the reservoir,
we consider an alternative and more practical measurement scheme by introducing
a side point-contact (PC) detector, as schematically shown in Fig.\ 1(a).
The PC-detector can perform continuous and {\it noninvasive} measurement
to reveal information about the electron's location in the dots or in the reservoir,
based on its distinct electrostatic effect on the tunnel barrier of the PC-detector.
We notice that the idea to replace the frequent projective measurements
by a genuine continuous measurement was also briefly discussed in Ref.\ \cite{Kur00}
in the context of verifying the anti-Zeno effect,
but no quantitative comparison was carried out there.
In this work we will carry out a quantitative comparison between
the continuous measurement by the PC-detector
and the frequent projective measurements.
The treatment employed in this work allows us to account for the non-Markovian
`return effect' from the reservoir in a rather transparent manner.
We also extend our analysis from the null-result conditioned evolution
to non-selective (ensemble averaged) dynamics,
which is proved to hold as well the desirable scaling property
and might be verified by the nowadays state-of-the-art experiments.

\begin{figure}[h]
\includegraphics[width=5.5cm]{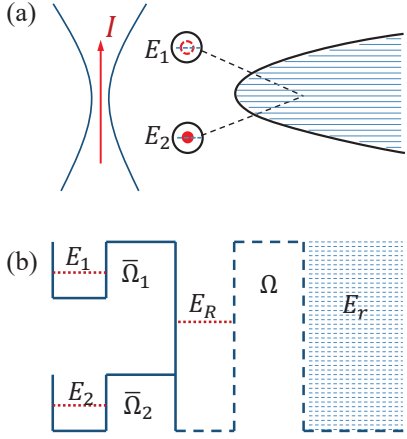}
\caption{
(a) Electron transfer through a continuum between two quantum dots
under continuous monitoring of a Point-Contact detector.
(b) The finite bandwidth (non-Markovian) continuum is equivalently described by a
fictitious well coupled to a fictitious infinite bandwidth (Markovian) reservoir. }
\label{fig1}
\end{figure}

\section{Model and Method}

Consider an electron in a double-dots system, where the two dots are
coupled in parallel to a continuum reservoir.
Moreover, the electron is monitored by a PC-detector, as shown in Fig.\ 1(a).
Let us as a first step neglect the PC detector.
The total Hamiltonian of the double dots coupled by the continuum is given by
\begin{align}\label{Hamorig}
&H=E_1|1\rangle\langle 1|+E_2|2\rangle\langle 2|+\sum_rE_r|r\rangle\langle r|\nonumber\\
&+\sum_r\Big[\big(\Omega_{1r} |r\rangle\langle 1|
+\Omega_{2r}|r\rangle\langle 2|\big)+{\rm h.c.}\Big] \,.
\end{align}
Here, $\Omega_{jr}$ is the coupling amplitude of the dot $j$ to the reservoir.
The states in the dots $|1(2)\rangle$ are localized
and the reservoir states $|r\rangle$ are continuum.

In this work, we are interested in a {\it finite bandwidth} continuum
which allows an {\it inverse motion}
of the electron from the continuum back into the dots,
i.e., a type of non-Markovian effect.
Following Ref.\ \cite{SG17}, an elegant {\it fictitious-well} model
can be employed to conveniently account for the non-Markovian effect.
One can imagine that the finite-band reservoir can be replaced by a localized state
(in a {\it fictitious well}) coupled to a wide-band continuum
of Markovian reservoir, as schematically shown in Fig.\ 1(b).
This consideration corresponds to separating the reservoir basis
$|r\rangle$ into two components \cite{SG17}
\begin{equation}\label{new-basis}
\sum_r |r\rangle\langle r|\to |R\rangle\langle R|
+\sum_{r^\prime} |r^\prime\rangle\langle r^\prime|  \,,
\end{equation}
where $|R\rangle$ denotes the localized state inside the fictitious well
and $|r'\rangle$ the extended states of a fictitious \emph{Markovian} reservoir
with constant density of states $\rho'$.
Then, the Hamiltonian in the new basis reads \cite{SG17}
\begin{align}\label{new-H}
&H=E_1|1\rangle\langle 1|+E_2|2\rangle\langle 2|+E_R|R\rangle\langle R|+\sum_{r^\prime}E_{r^\prime}|r^\prime\rangle\langle r^\prime|\nonumber\\
&~~~ +\sum_{j=1,2} (\bar{\Omega}_j|R\rangle\langle j|+{\rm H.c})
+\sum_{r^\prime}(\Omega|r^\prime\rangle\langle R|+{\rm H.c})  \,.
\end{align}
In order to make the two descriptions precisely equivalent,
the tunnel coupling amplitudes $\bar{\Omega}_{1,2}$ of the fictitious well
to the dots, and also to the Markovian reservoir, should be properly adjusted.
As proved in Appendix A, we should choose
\begin{align}
\bar\Omega_j=\sqrt{{\Gamma_j\Lambda\over2}} \,.
\end{align}
Here $\Gamma_j=2\pi\Omega^2_{jr}(E_R)\rho(E_R)$
are the coupling rates of the dot levels to the {\it original} finite-band
reservoir with density-of-states $\rho(E_R)$ at the spectral center ($E_R$),
and $\Lambda$ is the width of the reservoir spectrum.
Also, we should choose
\bea\label{Lam}
\pi\Omega^2\rho^\prime=\Lambda
\eea
for the coupling strength of the fictitious well state $|R\rangle$
to the Markovian reservoir (with density of states $\rho'$).
Under these choices, result from the {\it fictitious} Hamiltonian is precisely
the same as that from the original one, \Eq{Hamorig}.

Using the new basis, the time dependent state of the electron can be expressed as
\begin{align}\label{new-WF}
|\Psi (t)\rangle=b_1(t)|1\rangle +b_2(t)|2\rangle + b_R(t)|R\rangle
+\sum_{r'} b_{r'}(t)|r'\rangle \,.
\end{align}
Starting with the time dependent Scr\"odinger equation in the new basis,
we can first derive the equations for $\{b_1(t),b_2(t),b_R(t),b_{r'}(t)\}$.
Eliminating $b_{r'}(t)$, but keeping the state of the fictitious well,
we obtain equations for $\{b_1(t),b_2(t),b_R(t)\}$.
For the purpose that we will later introduce decoherence from,
e.g., the backaction of the PC-detector,
we employ here the description of density matrix with elements
$\rho_{i,j}(t)=b_i(t)b^*_j(t)$.
Here $i$ and $j$ denote, respectively, the basis states $\{1,2,R\}$.
We then obtain the following rate equations for the density matrix
\begin{subequations}
\label{Eqsigmavst}
\begin{align}
&\dot\rho_{11}=i\bar\Omega_1(\rho_{1R}-\rho_{R1}) \,,
\label{Eqsigmavsta}\\
&\dot\rho_{22}=i\bar\Omega_2(\rho_{2R}-\rho_{R2}) \,,
\label{Eqsigmavstb}\\
&\dot\rho_{RR}=i\bar\Omega_1(\rho_{R1}-\rho_{1R})
+i\bar\Omega_2(\rho_{R2}-\rho_{2R})
\nonumber\\
&~~~~~~~~~~~~~~~~~~~~~~~~~~~~~~~~~~~~~~~ -2\Lambda\rho_{RR} \,,
\label{Eqsigmavstc}\\
&\dot\rho_{12}=i(E_2-E_1)\rho_{12}+i(\bar\Omega_2\rho_{1R}
-\bar\Omega_1\rho_{R2})  \,,
\label{Eqsigmavstd}\\
&\dot\rho_{1R}=i(E_R-E_1)\rho_{1R}+i\bar\Omega_1(\rho_{11}-\rho_{RR}) \,,
\nonumber\\
&~~~~~~~~~~~~~~~~~~~~~~~~~~~~~+i\bar\Omega_2\rho_{12}-\Lambda\rho_{1R} \,,
\label{Eqsigmavste}\\
&\dot\rho_{2R}=i(E_R-E_2)\rho_{2R}+i\bar\Omega_2(\rho_{22}-\rho_{RR})
\nonumber\\
&~~~~~~~~~~~~~~~~~~~~~~~~~~~~~+i\bar\Omega_1\rho_{21}-\Lambda\rho_{2R} \,.
\label{Eqsigmavstf}
\end{align}
\end{subequations}
Based on theses equations, if we further
eliminate the degree of freedom of the fictitious well,
we obtain then the usual non-Markovian master equations,
with the non-Markovian nature reflected by a time-nonlocal memory form.
However, keeping in these equations the information of $|R\ra$,
the consequence is remarkable.
First, the non-Markovianity is manifested in this treatment,
quite physically, as a {\it return-back-effect} from the fictitious well.
This is something like the back-flow-of-information which has been
frequently discussed among the non-Markovian community.
We know that if the dots are connected directly with
a wide-band Markovian reservoir \cite{SG11,SG13,SG14},
the particle will never return back
once it has been confirmed (by measurement) in the reservoir.
However, in the non-Markovian case,
the particle can return back into the dots,
even if the particle has been confirmed in the well
but not in the wide-band reservoir.
This type of treatment clearly splits the `origin' of
the return-back-effect from the non-Markvian environment.

The second advantage of the fictitious-well treatment is allowing us
very easily to include the measurement back-action into the dynamics.
From Fig.\ 1 we understand that the measurement effect
of the PC detector is to generate dephasing
between the dots and the fictitious well states.
So we only need to insert dephasing terms
into Eqs.~(\ref{Eqsigmavst}d-f)
for the off-diagonal elements of the density matrix
\begin{subequations}
\label{EqPCvst}
\begin{align}
&\dot\rho_{12}=i(E_2-E_1)\rho_{12}+i(\bar\Omega_2\rho_{1R}
-\bar\Omega_1\rho_{R2})\nonumber\\
&~~~~~~~~~~~~~~~~~~~~~~~~~~~~~~~~
-\frac{(\sqrt{\Gamma_{d1}}-\sqrt{\Gamma_{d2}})^2}{2}\rho_{12} \,,\\
&\dot\rho_{1R}=i(E_R-E_1)\rho_{1R}+i\bar\Omega_1(\rho_{11}-\rho_{RR})
+i\bar\Omega_2\rho_{12}
\nonumber\\
&~~~~~~~~~~~~~~~~~~~~~~~~~~~~~~~~
-\left({\Gamma_{d1}\over2}+\Lambda\right)\rho_{1R} \,,
\label{EqPCvsta}\\
&\dot\rho_{2R}=i(E_R-E_2)\rho_{2R}+i\bar\Omega_2(\rho_{22}-\rho_{RR})
+i\bar\Omega_1\rho_{21}
\nonumber\\
&~~~~~~~~~~~~~~~~~~~~~~~~~~~~~~~~
-\left({\Gamma_{d2}\over2}+\Lambda\right)\rho_{2R} \,.
\label{EqPCvstb}
\end{align}
\end{subequations}
Here, for the sake of generality, we have considered the PC
detector unequally coupled to the two dots, thus
$\Gamma_{d1}\neq\Gamma_{d2}$.
This will take place if the detector is not set precisely
at the symmetric location with respect to the dots.

\section{Results and Discussions}

In Ref.\ \cite{SG14}, the transfer dynamics of an electron
through a finite-bandwidth non-Markovian reservoir was analyzed,
conditioned on {\it null-result} of measurement in the reservoir.
There, a perfect scaling behavior of the transfer dynamics
with the variable $x=\Lambda\tau$ was found,
see also Ref.\ \cite{Xu19} for arbitrary spectra.
Now, let us consider to replace the frequent {\it discrete} measurements
in the reservoir (we may imagine by an `external' observer),
by the {\it continuous} monitoring
using the more realistic PC-detector as shown in Fig.\ 1(a).
We would like to revisit this same problem and ask:
Can the $x=\Lambda\tau$-type scaling behavior still survive?
A key problem arising here is that, for the transfer dynamics
under continuous monitoring by the PC-detector,
it is almost impossible to realize the null-result conditioned transfer dynamics,
since the null-result of measurement excludes
the appearance of the electron in the reservoir.
Therefore, the {\it non-selective} transfer dynamics
under the continuous monitoring
is more natural, which corresponds to the statistical mixture
of the null-result and registered-result in the reservoir.
Note also that the non-Markovian {\it return effect} is to be included
in all the numerical simulations of this work,
either automatically for the continuous measurement by the PC-detector,
or applying the iterative algorithm outlined in Appendix B
for the frequent projective measurements.
In Appendix C, the return effect is explicitly displayed,
together with quantitative discussions.

\begin{figure}[H]
  \centering
  \includegraphics[scale=0.45]{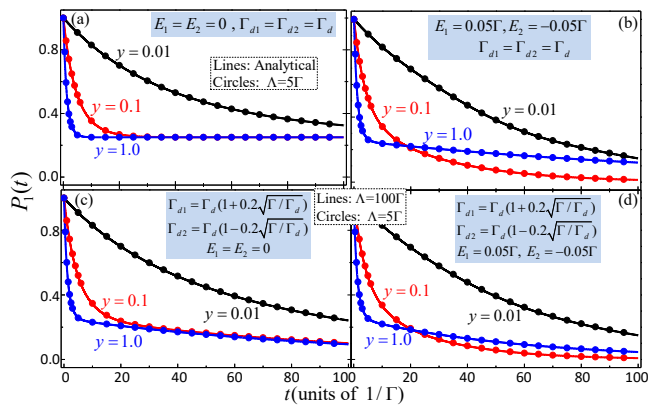}\\
  \caption{
  Probability of the electron remained in the initial quantum dot
  under continuous monitoring by a PC-detector.
  The demonstrating scaling property is characterized by the scaling variable
  $y=\Lambda/\Gamma_d$, with $\Lambda$ the bandwidth of the continuum
  and $\Gamma_d$ the (average) measurement rate.
  Set-up parameters:
  (a) $E_1=E_2=0$ and $\Gamma_{d1}=\Gamma_{d2}=\Gamma_d$;
  (b) $E_{1,2}=\pm 0.05\Gamma$ and $\Gamma_{d1}=\Gamma_{d2}=\Gamma_d$;
  (c) $E_1=E_2=0$ and $\Gamma_{d1,2}=\Gamma_d(1\pm 0.2\sqrt{\Gamma/\Gamma_d})$;
  (d) $E_{1,2}=\pm 0.05\Gamma$ and $\Gamma_{d1,2}=\Gamma_d(1\pm 0.2\sqrt{\Gamma/\Gamma_d})$.
    }\label{Figure04}
\end{figure}

For the double-dot setup shown in Fig.\ 1,
under the symmetric condition of $E_1=E_2=E_R$ and $\Gamma_{1}=\Gamma_{2}=\Gamma$,
one can obtain analytic result for the survival probability
of the electron in its initially occupied dot (i.e. in state $|1\ra$)
\begin{eqnarray}\label{P1t}
P_{1}(t)=\frac{1}{4}\left(e^{-\alpha\Gamma t}+1\right)^2 \,,
\end{eqnarray}
where
\begin{equation}\label{alphPC}
\alpha=\frac{2y}{1+2y}
~~~ {\rm with} ~~ y=\Lambda/\Gamma_d  \,.
\end{equation}
This result was obtained based on Ref.\ \cite{SG17},
by applying a basis transformation.
Importantly, we see that the result has an exact scaling behavior
with scaling variable $y=\Lambda/\Gamma_d$,
which resembles $x=\Lambda\tau$ associated with
the frequent measurements in previous studies \cite{SG11,SG13,SG14,Xu16,Xu18,Xu19}.

In the following numerical results,
we set $y=\Lambda/\Gamma_d$ as the scaling variable.
First, in Fig.\ 2(a), we display the results
for an ideal symmetric configuration,
i.e., $E_1=E_2=E_R$ and $\Gamma_{d1}=\Gamma_{d2}=\Gamma_d$.
We plot the numerical results of $P_1(t)$ for a couple of values
of the scaling parameter, say, $y=1$, 0.1, and 0.01
associated with a finite bandwidth $\Lambda=5\Gamma$.
We compare the results (labeled by symbols)
with the analytical solution of \Eqs{P1t} and (\ref{alphPC}) (solid lines).
This comparison is necessary by noting that the analytic result was
obtained under the limiting procedure $\Lambda\to\infty$ and
$\Gamma_d\to \infty$ for each given $y$.
Indeed, as shown in Fig.\ 2(a), perfect scaling behavior
survives in the {\it non-selective} dynamics
under the continuous monitoring of the PC detector.
This is an important addition to the previous studies
\cite{SG11,SG13,SG14,Xu16,Xu18,Xu19},
where both the artificial frequent measurements in the reservoir
and the restrictive null-result-conditioned dynamics were challengingly assumed.

Next, let us consider the effect of slight deviation from the ideal case.
In real set-up, the dot levels might be misaligned
and the PC detector may couple to the dots asymmetrically,
i.e., $E_1\neq E_2$ and/or $\Gamma_{d1}\neq \Gamma_{d2}$.
Notice that we are interested in the scaling property between $\Lambda$ and $\Gamma_d$.
We understand that the difference of $\Gamma_{d1}$ and $\Gamma_{d2}$
is, in general, not free from $\Gamma_d$
when we rescale $\Gamma_d$ with the change of $\Lambda$.
From Eq.\ (8a), we know that it is the difference of
$\sqrt{\Gamma_{d1}}$ and $\sqrt{\Gamma_{d2}}$
that determines the {\it deviation effect} (dynamics).
Let us denote this difference as
$\sqrt{\Gamma_{d1}}-\sqrt{\Gamma_{d2}}=\delta\sqrt{\Gamma}$,
with $\delta$ a small parameter,
and assume $\Gamma_{d1,2}=\Gamma_d\pm d$.
Under the condition $d<<\Gamma_d$,
we obtain $d=\delta\sqrt{\Gamma \Gamma_d}$.
Therefore, in the plot of Fig.\ 2(c) and (d), we apply this consideration
to characterize the difference of $\Gamma_{d1}$ and $\Gamma_{d2}$,
i.e., $\Gamma_{d1,2}=\Gamma_d(1 \pm \delta\sqrt{\Gamma/\Gamma_d})$.

The perfect coincidence between the dots and the lines in Fig.\ 2(b)-(d)
demonstrates indeed an overall scaling behavior.
Nevertheless, in practice, the difference of $\Gamma_{d1}$ and $\Gamma_{d2}$
may not satisfy the above requirement (i.e., $\propto \sqrt{\Gamma_d}$),
for instance, when we attempt to change the measurement rate
via altering the bias voltage ($V_d$) across the PC-detector.
This indicates that in experiments one should reduce or
eliminate the asymmetry leading to $\Gamma_{d1}\neq \Gamma_{d2}$.
For the results shown in Fig.\ 2, a common feature is that
with the increase of the measurement strength
(more precisely, for smaller $y$),
the decay of the initial occupation becomes slower.
This is nothing but the well-known Zeno effect.

More specifically,
in Fig.\ 2(b) we show the result for misaligned dot levels.
Unlike the case $E_1=E_2$ shown in Fig.\ 2(a),
here the electron will gradually immerse into the reservoir.
The basic reason is as follows.
For aligned levels ($E_1=E_2$), one can prove that,
by a simple basis transformation of the dot states,
$|\tilde{1}(\tilde{2})\ra=(|1\ra \mp |2\ra)/\sqrt{2}$,
the superposed bound state $|\tilde{1}\ra$
is isolated from the reservoir,
behaving like the `dark state' in quantum optics.
However, the decay dynamics under different $y$ in Fig.\ 2(b)
goes a little bit beyond our simple intuition,
e.g., a crossing of the blue and red curves for $y=1.0$ and $0.1$ takes place.

In Fig.\ 2(c) and (d), we show the results for $\Gamma_{d1}\neq \Gamma_{d2}$,
respectively, for aligned and misaligned dot levels.
Here, another physics is involved.
Owing to the decoherence between $|1\ra$ and $|2\ra$ caused by
the asymmetric measurement coupling ($\Gamma_{d1}\neq \Gamma_{d2}$),
the quantum-coherence-supported state $|\tilde{1}\ra$,
which is isolated from the reservoir, cannot be ideally formed.
Thus the electron will gradually leak into the reservoir
even for aligned dot levels, as shown in Fig.\ 2(c).   \\
\\

{\it Connection with Frequent Measurements}.---
Now we make a quantitative comparison
between the continuous measurement by the PC-detector
and the frequent projective measurements in the reservoir.
Actually, for the frequent measurements, one can obtain
the same result of $P_1(t)$ as \Eq{P1t},
but with $\alpha$ replaced by $\alpha^\prime$, given by \cite{SG17}
\begin{equation}\label{alphFM}
\alpha^\prime=1-(1-e^{-x})/x  \,.
\end{equation}
Here the scaling variable appears as $x=\Lambda\tau$,
with $\tau$ the time interval
between the frequent measurements in the reservoir.

This result, together with \Eqs{P1t} and (\ref{alphPC}),
reveals an interesting connection between
the two schemes of measurements.
Roughly speaking, as naively thought, $\Gamma^{-1}_d$
in the case of PC detector should correspond to
the time interval $\tau$ in the frequent measurements.
Indeed, if we identify $y=\Lambda/\Gamma_d=x$,
then $\alpha$ coincides with $\alpha^\prime$ for both $x\to 0$ and $x\to\infty$.
However, this identification does not hold for non-limiting regime.
A detailed comparison with $\alpha^\prime(x)$
shows that satisfactory agreement for $x>5$
can be achieved by inserting $\Gamma^{-1}_d=\tau/2$ into $\alpha(y)$,
while for small $x$ (e.g. $x<2$)
the identification $\Gamma^{-1}_d=\tau/4$ is better.
This latter identification can be analytically proved by
expanding $\alpha(y)$ and $\alpha^\prime(x)$
to the first order of the scaling parameters \cite{SG17}.   \\
\\

\begin{figure}[H]
  \centering
  \includegraphics[scale=0.6]{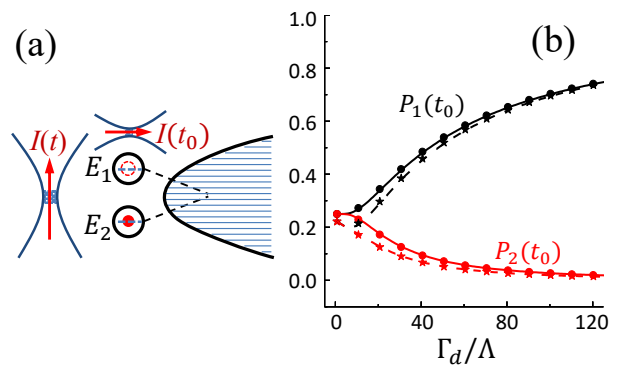}\\
  \caption{
Proposal for experimental demonstration.
(a) Schematic illustration of the proposed set-up:
in addition to the continuous monitoring by the left PC-detector
as shown in Fig.\ 1, a more PC-detector (the upper one)
is arranged to perform a fast projective measurement
for the occupation of the upper quantum dot,
by turning on it at a given time moment (e.g., at $t_0$).
(b)
Simulated results for the functional dependence of the upper and lower
quantum dots occupation probabilities
$P_{1,2}(t_0)|_{t_0=20\Gamma^{-1}}$
on the `inverse' scaling parameter $\Gamma_d/\Lambda$.
Notice that $\Gamma_d$ is proportional to the bias voltage $V_d$
across the continuous monitoring PC-detector.
This plot should make the predicted dependence
more closely related to the measurement data in experiments.
The results for the frequent measurements
are plotted by the solid and dashed lines for, respectively,
aligned dot levels ($E_1=E_2$)
and misaligned levels ($E_{1,2}=\pm 0.05\Gamma$),
while the results for the continuous measurement by the PC-detector
are displayed by the symbols (solid circles and stars).
In this perfect coincidence comparison, we have identified
the time interval by using $\tau=4\Gamma^{-1}_d$.  
All the results are obtained
under the choice of $\Lambda=5\Gamma$.   }
\end{figure}

{\it Proposal for Experimental Demonstration}.---
The transfer dynamics under continuous monitoring by the PC-detector,
especially the $y=\Lambda/\Gamma_d$ scaling property,
can be demonstrated by the state-of-the-art experiment.
By virtue of the high-quality fabrication and on-chip integration
of quantum dots and quantum-point-contacts,
we may propose the examination as schematically shown in Fig.\ 3(a).
In addition to the continuous monitoring by the left PC-detector
as shown in Fig.\ 1, a more PC-detector (the upper one)
is arranged here to perform a fast projective measurement
for the occupation probability of the electron in the upper quantum dot,
by turning on it at a given moment, e.g., at $t_0$.

Under continuous monitoring, the transfer dynamics is fully characterized
by the occupation probabilities $P_{1,2}(t)$ of the quantum dots.
For instance, in the special case of $E_1=E_2$,
$P_1(t)$ is analytically given by \Eqs{P1t} and (\ref{alphPC}),
which clearly displays the {\it functional dependence}
on the scaling variable $y=\Lambda/\Gamma_d$.
To make the dependence more experimentally relevant,
we may plot against the bias voltage ($V_d$) of the PC detector,
since the measurement rate $\Gamma_d$ is related to $V_d$
simply via $\Gamma_d=V_d (\sqrt{T}-\sqrt{T'})^2/2\pi$.
Here $T$ and $T'$ are the respective tunneling amplitudes of electron
through the point-contact, conditioned on the double-dots occupied or not.
Therefore, a $V_d$ dependence (in experiment)
is equivalent to the $\Gamma_d$ dependence,
or, the `inverse' scaling variable dependence
by noting that $y^{-1}=\Gamma_d/\Lambda$.

In Fig.\ 3(b), for the special case $E_1=E_2$,
we plot $P_{1,2}(t_0)|_{t_0=20\Gamma^{-1}}$
against $\Gamma_d/\Lambda$, by the black and red dots.
Also, we plot the result for the frequent measurement
by the solid black and red lines,
after making the identification $\tau=4\Gamma^{-1}_d$.
The perfect coincidence between the dots and lines shown here
supports as well our above discussion and conclusion
on the connection between the {\it frequent} and {\it continuous} measurements.
In Fig.\ 3(b), we also plot the results obtained numerically
for the misaligned levels ($E_{1,2}=\pm 0.05\Gamma$),
under both the continuous and frequent measurements.

We may further discuss the feasibility of possible experimental investigations.
Actually, sensitive detectors capable of fast, high-fidelity, single shot
measurements of quantum states are essential to solid-state quantum computation,
in particular to the readout of semiconductor-based charge and/or spin qubits.
In practice, the quantum-point-contact charge detector has been demonstrated
and employed by a large number of experiments,
e.g., in Refs.\ \cite{Ens06,Fuji06,Guo13,Guo15,Guo16}.
Usually, the read-out time is longer than 100 $\mu$s,
which is limited by the measurement bandwidth of the sensor current
(e.g., a high-frequency cutoff of typically less than 100 kHz).
However, in recent years, great efforts have been made to improve the readout speed
by employing the idea of rf reflectometry measurements.
That is, integrate the point-contact sensor into an rf microwave resonator circuit
and the reflectometry signal is highly sensitive to charge configurations.
The latest progress of this technique has realized
the total linewidth of $\sim 2$ MHz,
which sets the maximum measurement bandwidth
and reaches the time resolution about 1 $\mu$s \cite{Taru19,Kei19,Van19}.

For our purpose to demonstrate the proposal schematically shown in Fig.\ 3,
we would like to say that the requirement can be lower
than the semiconductor-based qubit measurements.
In our system, the narrow-bandwidth continuum can be realized by coupling
a single-level quantum dot to a  wide-band Markovian reservoir,
just as shown in Fig.\ 1 by the fictitious-well model.
The coupling strength between this quantum dot and the wide-band reservoir
is tunable, which determines the bandwidth ($\Lambda$)
of the non-Markovian reservoir, see \Eq{Lam}.
Then, even for the normal readout rate of PC detector,
e.g., $\Gamma^{-1}_d= 10\sim 100$ $\mu$s,
one can accordingly tune the coupling
to make the bandwidth $\Lambda$ of the same magnitude of $\Gamma_d$,
which is required to make the continuous monitoring effect prominent.
By virtue of this tunability of the coupling strength (thus of $\Lambda$),
one can even more directly demonstrate the scaling behavior as shown in Fig.\ 2,
rather than the less direct way as by Fig.\ 3(b).
Moreover, concerning the strong projective measurement
by the second PC detector (the upper one in Fig.\ 3(a)),
we can say that the readout is also not difficult.
One can design weak coupling of the double dots
to the reservoir, i.e., a small $\Gamma$ such as $\Gamma^{-1}$
on a timescale of  millisecond.
This implies a slow leaky and transfer process,
compared with the projective measurement by the second PC detector.

\section{Concluding Remarks}

The conceptual picture of frequent projective measurements performed in the reservoir
was typically employed in the studies of Zeno effect and quantum trajectories.
However, how the frequent measurements in the reservoir
are replaced by continuous measurement
performed by an {\it external} realistic detector/sensor
is a very interesting problem, as briefly discussed in Ref.\ \cite{Kur00}.  
In this work, taking the specific setup of electron transfer
through a continuum between two quantum dots as an example,
we have presented a study which is relevant to this issue.
The continuous monitoring is implemented by a point-contact detector.
It was found that the continuous measurement rate ($\Gamma_d$) is indeed
related to the time interval ($\tau$) of the frequent measurements
qualitatively as $\Gamma_d \simeq \tau^{-1}$,
yet which is quantitatively valid only in limiting regimes
and needs modification by multiplying a proper
proportional coefficient (e.g., 4 or 2) in non-limiting regimes.

We have considered the continuum with a finite bandwidth ($\Lambda$)
and employed a fictitious-well model to account for the non-Markovian
`return effect' in a transparent manner.
We also extended our analysis from the null-result conditioned evolution
to non-selective (ensemble averaged) dynamics
which is proved to hold as well the desirable $y=\Lambda/\Gamma_d$-scaling property
and is expected to verify by the state-of-the-art experiments.

\vspace{0.3cm}
{\flushleft\it Acknowledgements.}---
We thank Shmuel Gurvitz for stimulating discussions
and especially introducing the fictitious-well model to us.
We also thank Gang Cao and Guo-Ping Guo for valuable communications
on the experimental issues of PC-detector measurements.
This work was supported by the
National Key Research and Development Program of China
(No.\ 2017YFA0303304) and the NNSF of China (Nos.\ 11675016 \& 11974011\&11904261).

\appendix
\section{Fictitious-Well Model}

In this Appendix we review the derivation of the {\it fictitious-well} model,
proposed originally in Ref.\ \cite{SG17} for a single dot coupled to a continuum.
This model provides also an efficient description
for the setup of double dots as shown in Fig.\ 1(a).
In the `natural' basis,
the system (not including the point-contact detector)
is described by the Hamiltonian of \Eq{Hamorig}.
The electron's motion is described by the wavefunction
\begin{align}\label{WF-1}
|\Psi (t)\rangle=b_1(t)|1\rangle +b_2(t)|2\rangle +\sum_r b_{r}(t)|r\rangle \,.
\end{align}
Substituting this wavefunction into the time-dependent Schr\"{o}dinger equation
$i\partial_t|\Psi(t)\rangle=H|\Psi(t)\rangle$, we obtain
a set of differential equations for $b_{1,2}(t)$ and $b_r(t)$.
Applying the Laplace transformation and eliminating $b_r$, we obtain
\begin{subequations}\label{eqnormb1b2}
\begin{eqnarray}
(\omega-E_1)\tilde{b}_1(\omega)-
\sum_r\Omega_{1r}\frac{\Omega_{1r}\tilde{b}_1(\omega)
+\Omega_{2r}\tilde{b}_2(\omega)}{\omega-E_r}=ib_1(0) \,, \nonumber\\
\\
(\omega-E_2)\tilde{b}_2(\omega)-
\sum_r\Omega_{2r}\frac{\Omega_{1r}\tilde{b}_1(\omega)
+\Omega_{2r}\tilde{b}_2(\omega)}{\omega-E_r}=ib_2(0) \,, \nonumber\\
\end{eqnarray}
\end{subequations}
where $\tilde{b}_j(\omega)=\int_0^\infty b_j(t)(t)e^{i\omega t}dt$,
with $j=1,2$ and $\omega\to \omega+i0^+$.
In this result,
$b_{1,2}(0)$ are the initial amplitudes of the electron in the double dots.

Let us assume a Lorentzian spectral density function
for the finite bandwidth reservoir
\begin{align}
\Omega^{}_{jr}\Omega^{}_{j'r}\rho (E_r)
={\Lambda^2\sqrt{\Gamma_{j}\Gamma_{j'}}\over 2\pi[(E_r-E_R)^2+\Lambda^2]} \,,
\label{lor}
\end{align}
where $\Lambda$ is the bandwidth of the spectrum,
and $\Gamma_{j}^{}=2\pi\Omega^{2}_{j}(E_R)\rho(E_R)$
characterize the coupling strengths for $j,j'=\{1,2\}$.
Here we drop the label `$r$' from $\Omega_{jr}$,
indicating an overall coupling strength.
The level $E_R$ corresponds to the Lorentzian center,
and the density-of-states $\rho(E_r)$ is introduced to replace
$\sum_r\rightarrow \int\rho(E_r)dE_r$.
Then we carry out the integration and obtain
\begin{equation}\label{Fjj}
\int {\Omega^{}_{jr}\Omega^{}_{j'r}\over \omega-E_r} \rho(E_r)dE_r
={\Lambda \sqrt{\Gamma_j\Gamma_{j'}}\over 2(\omega-E_R+i\Lambda)}  \,.
\end{equation}
Now let us introduce an auxiliary amplitude
\bea
\tilde{b}_R(\omega)={\bar{\Omega}_1\tilde{b}_1(\omega)
+\bar{\Omega}_2\tilde{b}_2(\omega)\over \omega-E_R+i\Lambda}  \,,
\eea
where, in particular, we set
\bea
\bar{\Omega}_{1,2}=\sqrt{\Gamma_{1,2}\Lambda \over 2}  \,.
\eea
Under this construction, one can prove that the following equations
precisely recover the original result of Eqs.\ (\ref{eqnormb1b2})
\begin{subequations}\label{EqNBab1b2}
\begin{align}
&(\omega-E_j)\tilde{b}_j(\omega)-\bar{\Omega}_j\tilde{b}_R(\omega)=ib_j(0) \,,
\\
&(\omega-E_R+i\Lambda)\tilde{b}_R(\omega)
-\sum_{j=1,2}\bar{\Omega}_j\tilde{b}_j(\omega)=0  \,.
\end{align}
\end{subequations}
Physically, this set of equations corresponds to the {\it fictitious well} model
depicted by Fig.\ 1(b), where the non-Markovian component
(i.e. the {\it fictitious well}) is extracted out
from the extended continuum $|r\rangle$ of the reservoir.

\section{Frequent Measurements in the Reservoir}

For a finite-bandwidth reservoir,
one of the non-Markovian consequences is a {\it return-effect}
of the particle from the reservoir to the system (here, the double dots).
For the continuous monitoring by the PC-detector,
the non-selective dynamics shown in the main text
has automatically contained the return-effect.
In this Appendix, we present the explicit treatment for the return-effect in the
non-selective dynamics associated with frequent measurements in the reservoir.
This non-trivial procedure has been involved in
carrying out the results in Fig.\ 3(b).

The main idea is developing an {\it iterative approach}
to the `evolution-plus-measurement' dynamics.
For each time interval $\tau$ (note that $t=n\tau$),
the particle is subject first to
a free evolution described by \Eq{Eqsigmavst},
then to a projective measurement with result
either in the double-dots or in the fictitious well.
Note that the possible result in the effective wide-band reservoir ($\{|r'\ra\}$)
has been ruled out in \Eq{Eqsigmavst}.

The free evolution is described by \Eq{Eqsigmavst},
where the truncated/projected state (density matrix) is defined by
eliminating the components in the wide-band reservoir ($\{|r'\ra\}$)
from the wave-function of \Eq{new-WF}.
That is, the truncated state is described by the
$3\times3$ density matrix $\rho(t)$
with elements $\rho_{ij}(t)=b_i(t)b_j^*(t)$,
while the index $i(j)=1$, $2$, and $R$
corresponds to the electron in the dots and the fictitious well, respectively.
Formally, we describe the evolution as
\begin{equation}
\rho(t+\tau)={\cal U}(\tau)\rho_M(t){\cal U}^\dag(\tau) \,,
\end{equation}
where $\rho_M(t)$ is the statistical mixture
by averaging the measurement on the state $\rho(t)$.
This will be clear after we determine the result of $\rho_M(t+\tau)$.

Now, based on $\rho(t+\tau)$, let us introduce a measurement on it.
If the electron is found in the dots (i.e., not in the well),
the measurement Kraus operator can be expressed
(in the basis $\{|1\ra, |2\ra, |R\ra \}$) as
$\mathcal{M}_0=diag\{1,1,0\}$,
and the resultant state reads
\begin{equation}
\rho_0(t+\tau)=\mathcal{M}_{0}\rho(t+\tau)\mathcal{M}_{0}^\dag/||\bullet|| \,,
\end{equation}
where $||\bullet||$ denotes the normalization factor.
Note that the probability of finding the result is right
this normalization factor, say,
$P_0=\rm{Tr}[{\cal E}_0\rho(t+\tau)]$,
where the POVM operator is given by
${\cal E}_0=\mathcal{M}_{0}^\dag\mathcal{M}_{0}$.
Similarly, if the electron is found in the well,
the measurement Kraus operator is
$\mathcal{M}_R=diag\{0,0,1\}$ and the resultant state reads
\begin{equation}
\rho_R(t+\tau)=\mathcal{M}_{R}\rho(t+\tau)\mathcal{M}_{R}^\dag /||\bullet|| \,.
\end{equation}
Also, here $||\bullet||$ denotes the normalization factor
and the probability of finding the electron in the well
is given by $P_R=\rm{Tr}[{\cal E}_R\rho(t+\tau)]$,
with ${\cal E}_R=\mathcal{M}_{R}^\dag\mathcal{M}_{R}$.
Finally, the non-selective dynamics is given by the statistical mixture
of the above two results:
\bea
&&\rho_M(t+\tau)
= P_0\,\rho_0(t+\tau) + P_R\,\rho_R(t+\tau)  \nl
&&~~ =\mathcal{M}_{0}\rho(t+\tau)\mathcal{M}_{0}^\dag
+\mathcal{M}_{R}\rho(t+\tau)\mathcal{M}_{R}^\dag   \,.
\eea

We then obtain the {\it iteration rule} for, respectively,
the measurement-result-conditioned states
and their statistical mixture as follows:
\begin{subequations}
\begin{align}
\rho_0(n)&={\cal U}_0(\tau)[\rho_0(n-1)+\rho_R(n-1)]{\cal U}^\dag_0(\tau) \,, \\
\rho_R(n)&={\cal U}_R(\tau)[\rho_0(n-1)+\rho_R(n-1)]{\cal U}^\dag_R(\tau) \,, \\
\rho_M(n)&={\cal U}_0(\tau)\rho_M(n-1){\cal U}^\dag_0(\tau) \nl
& ~~ + {\cal U}_R(\tau)\rho_M(n-1){\cal U}^\dag_R(\tau)  \,.
\end{align}
\end{subequations}
Here we have introduced the joint-operator
${\cal U}_{0,R}(\tau)=\mathcal{M}_{0,R}{\cal U}(\tau)$,
and the abbreviation $n$ and $n-1$ to denote
the states at the moments $t=n\tau$ and $(n-1)\tau$.
Using this method, one can straightforwardly carry out
the numerical results for generic configurations of the set-up.

\section{Illustration of Return Effect}

In this Appendix, in addition to the iterative algorithm outlined in Appendix B,
we more explicitly show the return effect.
From a simple intuition, a narrower bandwidth reservoir will result in
stronger non-Markovian effect, thus a stronger `return effect'
or the so called `back-flow-of-information' effect.
Another simple point is that, for a given bandwidth $\Lambda$,
longer time $\tau$ will result in stronger return effect.
Indeed, this expected behavior is demonstrated in Fig.\ 4(a)
by the first stage of time evolution (before reaching the maximum probability).
Note that the decay after the maximum peak in the next stage
is due to the electron gradually leaking into the attached wide-band reservoir.
In Fig.\ 4(b), we further check an example
that the electron is initially in the dots, e.g., in the upper dot.

More quantitatively, let us consider the most relevant parameters
$\Lambda=5\Gamma$ and $y=1$ in Fig.\ 2,  which correspond to
the narrowest bandwidth and the longest time ($\tau=0.2 \Gamma^{-1}$) shown there.
For them, from Fig.\ 4(a), we identify $P_1\simeq 0.037$, which indicates
a probability percentage of $7.4\%$ returning back to the double dots,
once the electron is registered in the fictitious well.
Similarly, for the results shown in Fig.\ 3(b) where the bandwidth is fixed
as $\Lambda=5\Gamma$,
another most relevant parameter is $\tau\simeq 0.43 \Gamma^{-1}$,
which roughly corresponds to the lateral coordinate
$\Gamma_d/\Lambda\simeq 1.86$.
This parameter (the value of $\tau$) is actually determined by the peak position
of the curve associated with $\Lambda=5\Gamma$ in Fig.\ 4,
while the height of the peak is $P_1\simeq 0.058$.
Therefore, this indicates an even larger percentage ($11.6\%$)
of returning-back probability for the electron registered in the fictitious well.

\begin{figure}[H]
  \centering
  \includegraphics[scale=0.65]{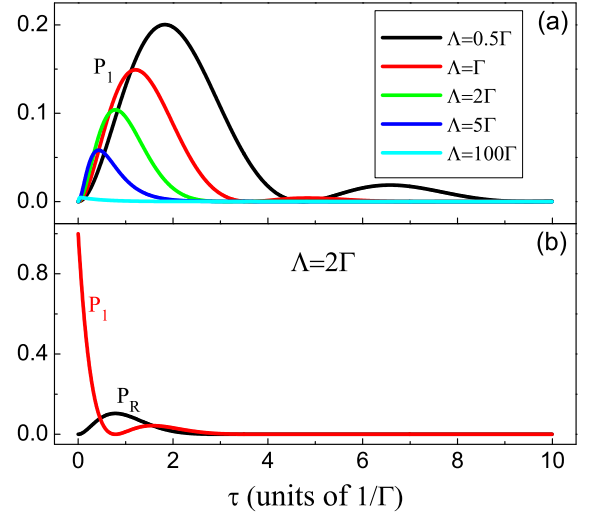}\\
  \caption{
Return effect from finite-bandwidth reservoirs.
(a)
For a couple of bandwidths and assuming the electron initially in the fictitious well,
the probability of appearing in the upper dot ($P_1$) as a function of time ($\tau$).
(b)
Assuming the electron initially in the upper dot,
the probabilities of remaining in this same dot ($P_1$)
and appearing in the fictitious well ($P_R$).  }
\end{figure}

\end{CJK*}
\end{document}